\documentclass[twocolumn,showpacs,preprintnumbers,amsmath,amssymb, prl]{revtex4}

\usepackage{graphicx}
\usepackage{dcolumn}
\usepackage{bm}

\begin{document}

\preprint{APS/123-QED}

\title{Dynamics Analysis Of Plasmon Resonance Modes In Nanoparticles}

\author{I. D. Mayergoyz, Z. Zhang}
\affiliation{Department of Electrical and Computer Engineering, Institute for Advanced Computer Studies, University of Maryland,
College Park, Maryland, 20742, USA}

\author{G. Miano}
\affiliation{Univ Naples Federico II, Dept Elect Engn, Via Claudio 21, Naples, I-80125 Italy}

\date{\today}

\begin{abstract}
A novel theoretical approach to the dynamics analysis of excitation of plasmon modes in nanoparticles is presented. This
approach is based on the biorthogonal plasmon mode expansion and it leads to the predictions of time-dynamics of excitation of
specific plasmon modes as well as their steady state amplitude and their decay. Temporal characteristics of plasmon modes in
nanoparticles are expressed in terms of their shapes, permittivity dispersion relations and excitation conditions. In the case
of the Drude model, analytical expressions for time-dynamics of plasmon modes are obtained.
\end{abstract}

\maketitle

Plasmon resonances in nanoparticles have been the focus of considerable experimental and theoretical research lately. This
research is motivated by numerous scientific and technological applications of these resonances in such areas as near-field
microscopy, nano-lithography, surface enhanced Raman scattering, nanophotonics, biosensors, optical data storage, etc. Steady
state properties of plasmon resonances in nanoparticles have been mostly studied, while the dynamics of specific plasmon modes
is the least understood area of plasmonics. This state of affairs has prompted the current burst of activity in the experimental
research of plasmon dynamics. This research is exemplified by publications
\cite{Lamprecht99,Lehmann00,Kubo05,Muskens06,Zentgraf04,Gay06,Lamprecht97}, where the plasmon dynamics and plasmon decay
(dephasing) rates have been extensively studied by using advanced femtosecond techniques. The theoretical temporal analysis of
plasmon modes is also very important to fully comprehend the time-dynamics of mode excitation and decay as well as to estimate
their steady state amplitude. The temporal analysis of plasmon resonance modes can be very instrumental in the area of light
controllability of plasmon resonances in semiconductor nanoparticles\cite{Fredkin2003,Mayergoyz2005}, where proper time
synchronizations of excitations of specific plasmon modes may be needed.

In this Letter, we present a self-consistent theoretical technique for the temporal analysis of specific plasmon modes in
nanoparticles and compare (where it is possible) the theoretical results with experimental data. This technique is based on the
biorthogonal plasmon mode expansion and it leads to the predictions of time-dynamics of excitation of specific plasmon modes and
their decay as well as their steady state amplitude in terms of nanoparticle shapes, permittivity dispersion relations and
excitation conditions. For instance, an explicit formula is derived for the steady state amplitude of plasmon modes in terms of
real and imaginary parts of dielectric permittivity, amplitude of incident field and its spatial orientation with respect to
dipole moments of the plasmon modes. In the case of the Drude model, analytical expressions for plasmon dynamics are obtained
which suggest that the reciprocal of the Drude damping factor can be identified with the decay (dephasing) time of plasmon
modes.

To start the discussion, consider a nanoparticle with boundary $S$ and dielectric permittivity $\epsilon(\omega)$ in free-space
with dielectric constant $\epsilon_{0}$. It has been demonstrated \cite{Fredkin2003,Mayergoyz2005,Ouyang891,Ouyang892} that the
resonance values $\epsilon_{k}$ of dielectric permittivity and corresponding resonant plasmon modes can be found by solving the
eigenvalue problems for the boundary integral equation
\begin{equation}
\sigma_{k}\left(Q\right)=\frac{\lambda_{k}}{2\pi}\oint_{S}\sigma_{k}\left(M\right)
\frac{\textbf{r}_{MQ}\cdot\textbf{n}_{Q}}{r_{MQ}^{3}}dS_{M},\label{eq:IE1}
\end{equation}
or its adjoint equation
\begin{equation}
\tau_{k}\left(Q\right)=\frac{\lambda_{k}}{2\pi}\oint_{S}\tau_{k}\left(M\right)
\frac{\textbf{r}_{QM}\cdot\textbf{n}_{M}}{r_{QM}^{3}}dS_{M},\label{eq:IE2}
\end{equation}
where $\sigma_{k}(M)$ has the physical meaning of surface electric charges on $S$ that produce electric field $\mathbf{E}_{k}$
of $k$-th plasmon mode, $\tau_{k}(M)$ has the physical meaning of dipole densities on $S$ that produce displacement field
$\mathbf{D}_{k}$ of $k$-th plasmon mode, while all other notations have their usual meaning. The resonance values of
$\epsilon_{k}$ and $\lambda_{k}$ are related by
\begin{equation}
\lambda_{k}=\frac{\epsilon_{k}-\epsilon_{0}}{\epsilon_{k}+\epsilon_{0}}.\label{eq:Eq3}
\end{equation}

After $\lambda_{k}$ are computed and resonance values of permittivity $\epsilon_{k}$ are found by using (\ref{eq:Eq3}),
resonance frequencies are determined from the dispersion relation:
\begin{equation}
\epsilon_{k}=\epsilon'(\omega_{k})=Re\left[\epsilon(\omega_{k}) \right].\label{eq:Eq4}
\end{equation}

It turns out that eigenfunctions $\sigma_{k}(M)$ and $\tau_{i}(M)$ are biorthogonal:
\begin{equation}
\oint_{S}\sigma_{k}(M)\tau_{i}(M)dS_{M}=\delta_{ki}.\label{eq:Eq5}
\end{equation}
Thus, the set of eigenfunctions $\sigma_{k}(M)$ can be used for the biorthogonal expansion of actual boundary charges
$\sigma(M,t)$ induced on particle boundary during the excitation process:
\begin{equation}
\sigma(M,t)=\sum_{k=1}^{\infty}a_{k}(t)\sigma_{k}(M),\label{eq:Eq6}
\end{equation}
where, according to (\ref{eq:Eq5}), the expansion coefficient $a_{k}(t)$ is given by the formula:
\begin{equation}
a_{k}(t)=\oint_{S}\sigma(M,t)\tau_{k}(M)dS_{M}.\label{eq:Eq7}
\end{equation}
It is clear that the time evolution of the expansion coefficient $a_{k}(t)$ reveals the time-dynamics of $k$-th plasmon mode
corresponding to the eigenfunction $\sigma_{k}(M)$. Since the medium of resonant nanoparticle is dispersive and exhibits
nonlocal in time constitutive relation $\mathbf{D}(t)$ vs $\mathbf{E}(t)$, the frequency domain technique will be employed for
the calculations of $a_{k}(t)$. This means that the equation (\ref{eq:Eq6})-(\ref{eq:Eq7}) will be Fourier transformed
\begin{equation}
\tilde{\sigma}(M,\omega)=\sum_{k=1}^{\infty}\tilde{a}_{k}(\omega)\sigma_{k}(M),\label{eq:Eq8}
\end{equation}
\begin{equation}
\tilde{a}_{k}(\omega)=\oint_{S}\tilde{\sigma}(M,\omega)\tau_{k}(M)dS_{M},\label{eq:Eq9}
\end{equation}
and the equation for $\tilde{a}_{k}(\omega)$ will be first derived and solved. Subsequently, $a_{k}(t)$ will be found through
inverse Fourier transform.

To derive the equation for $\tilde{a}_{k}(\omega)$, the Fourier transformed boundary condition for the normal components of
electric field on $S$ is invoked:
\begin{equation}
\!\epsilon(\omega)\tilde{E}_{n}^{+}(Q,\omega)\!-\!\epsilon_{0}\tilde{E}_{n}^{-}(Q,\omega)=\left[\epsilon_{0}\!-\!\epsilon(\omega)
\right]\tilde{E}_{n}^{(0)}(Q,\omega).\label{eq:Eq10}
\end{equation}
Here: $\tilde{E}_{n}^{+}(Q,\omega)$ and $\tilde{E}_{n}^{-}(Q,\omega)$ are the limiting values of the normal components of
Fourier transformed electric field at $Q \in S$ from inside and outside $S$, respectively, while $\tilde{E}_{n}^{(0)}(Q,\omega)$
is the Fourier transformed normal component of the incident field on $S$.

By using formula $\tilde{\sigma}(Q,\omega)=\epsilon_{0}[\tilde{E}_{n}^{-}(Q,\omega)-\tilde{E}_{n}^{+}(Q,\omega)]$, the boundary
condition (\ref{eq:Eq10}) can be rearranged as follows:
\begin{equation}
\frac{\tilde{\sigma}(Q,\omega)}{\epsilon(\omega)-\epsilon_{0}}-\tilde{E}_{n}^{+}(Q,\omega)=\tilde{E}_{n}^{(0)}(\omega).\label{eq:Eq11}
\end{equation}
Now it can be recalled \cite{Kellogg29,Mikhlin70} that
\begin{equation}
\tilde{E}_{n}^{+}(Q,\omega)=-\frac{\tilde{\sigma}(Q,\omega)}{2\epsilon_{0}}+\frac{1}{4\pi
\epsilon_{0}}\oint_{S}\tilde{\sigma}(Q,\omega)\frac{\mathbf{r}_{MQ}\cdot \mathbf{n}_{Q}}{r_{MQ}^{3}}dS_{M}.\label{eq:Eq12}
\end{equation}

By substituting (\ref{eq:Eq12}) into (\ref{eq:Eq11}) and then by using expansion (\ref{eq:Eq8}) and formulas (\ref{eq:IE1}) and
(\ref{eq:Eq9}), the following expression for $\tilde{a}_{k}(\omega)$ is derived:
\begin{equation}
\tilde{a}_{k}(\omega)= \frac{2\epsilon_{0} \lambda_{k} \left[\epsilon(\omega)-\epsilon_{0} \right ]}{2\epsilon_{0} \lambda_{k}\!
+\! \left[\epsilon(\omega)\!-\!\epsilon_{0} \right
](\lambda_{k}\!-\!1)}\!\oint_{S}\!\tilde{E}_{n}^{(0)}(Q,\omega)\tau_{k}(Q)dS_{Q}.\label{eq:Eq13}
\end{equation}
The last integral can be simplified, because the incident field $\mathbf{E}^{(0)}(Q,t)$ is practically uniform within
nanoparticles. Thus, this field can be represented as $\mathbf{E}^{(0)}(Q,t)=\mathbf{E}_{0}f(t)$. By using this form of
$\mathbf{E}^{(0)}(Q,t)$ in (\ref{eq:Eq13}) and taking into account formula (\ref{eq:Eq3}) and the fact that due to the
``scaling'' freedom provided by normalization condition (\ref{eq:Eq5}) the following expression
$\oint_{S}\mathbf{n}_{Q}\tau_{k}(Q)dS_{Q}=(\epsilon_{k}-\epsilon_{0})\mathbf{p}_{k}$ for the dipole moment $\mathbf{p}_{k}$ of
$k$-th plasmon mode (see \cite{Mayergoyz2005}) can be used, the last formula can be represented in the form:
\begin{equation}
\tilde{a}_{k}(\omega)=(\mathbf{E}_{0}\cdot \mathbf{p}_{k})\tilde{g}_{k}(\omega)\tilde{f}(\omega),\label{eq:Eq14}
\end{equation}
where
\begin{equation}
\tilde{g}_{k}(\omega)=\frac{\epsilon(\omega)-\epsilon_{0}}{\epsilon_{k}-\epsilon(\omega)},\label{eq:Eq15}
\end{equation}
and $\tilde{f}(\omega)$ is the Fourier transform of $f(t)$.

Formulas (\ref{eq:Eq14})-(\ref{eq:Eq15}) are remarkably simple and they clearly suggest that $\tilde{g}_{k}(\omega)$ can be
construed as normalized (by $\mathbf{E}_{0}\cdot \mathbf{p}_{k}$) transfer function of $k$-th plasmon mode. The formula
(\ref{eq:Eq15}) reveals the resonance nature of excitation of $k$-th plasmon mode at the frequency $\omega_{k}$. Indeed,
according to (\ref{eq:Eq4}), at this frequency $\epsilon_{k}-\epsilon(\omega_{k})=-iIm\left[ \epsilon(\omega_{k})\right]$ and
the magnitude of $\tilde{g}_{k}(\omega)$ will be narrow peaked if $\epsilon''(\omega_{k})$ is sufficiently small, that is when
the specific plasmon resonance is strongly pronounced.

By using formula (\ref{eq:Eq14}), the following expression is derived for $a_{k}(t)$:
\begin{equation}
a_{k}(t)=(\mathbf{E}_{0}\cdot \mathbf{p}_{k})\int_{0}^{t}g_{k}(t-t')f(t')dt',\label{eq:Eq16}
\end{equation}
with $g_{k}(t)$ being the inverse Fourier transform of $\tilde{g}_{k}(\omega)$. Thus, the algorithm of analysis of time-dynamics
of $k$-th plasmon mode can be stated as follows. First, the eigenvalue problem (\ref{eq:IE1}) and (\ref{eq:Eq3}) is solved and
resonance values $\epsilon_{k}$ of dielectric permittivity and surface electric charges $\sigma_{k}(M)$ of the corresponding
plasmon mode are found. Next, by using $\sigma_{k}(M)$, the dipole moment $\mathbf{p}_{k}$ is computed. Then, $g_{k}(t)$ is
determined through the inverse Fourier transform of $\tilde{g}_{k}(\omega)$ given by (\ref{eq:Eq14}). Finally, formula
(\ref{eq:Eq16}) is employed to evaluate the time evolution of $a_{k}(t)$ which reveals the time-dynamics of $k$-th plasmon mode.
It is worthwhile to mention that the outlined computations can be performed by using actual, experimentally measured
$\epsilon(\omega)$.

Expressions (\ref{eq:Eq14})-(\ref{eq:Eq16}) can be useful for analytical calculations as well. To demonstrate this, consider the
steady state case when $f(t)=sin\omega_{k}t$, where $\omega_{k}$ is the resonance frequency defined by the formula
(\ref{eq:Eq4}). Then, $\tilde{f}(\omega)=i \sqrt{\frac{\pi}{2}}\left[
\delta(\omega-\omega_{k})-\delta(\omega+\omega_{k})\right]$ and, by using (\ref{eq:Eq14})-(\ref{eq:Eq15}) as well as the inverse
Fourier transform of $\tilde{a}_{k}(\omega)$, we arrive at:
\begin{equation}
a_{k}^{(ss)}(t)=-(\mathbf{E}_{0}\cdot \mathbf{p}_{k})\left[
\frac{\epsilon'(\omega_{k})-\epsilon_{0}}{\epsilon''(\omega_{k})}cos\omega_{k}t + sin\omega_{k}t \right],\label{eq:Eq17}
\end{equation}
where superscript $(ss)$ indicates the steady state of $a_{k}(t)$.

In the case of strong plasmon resonances when $|\epsilon'(\omega_{k})-\epsilon_{0}|>>|\epsilon''(\omega_{k})|$, the last formula
is simplified as follows:
\begin{equation}
a_{k}^{(ss)}(t)=-(\mathbf{E}_{0}\cdot \mathbf{p}_{k})
\frac{\epsilon'(\omega_{k})-\epsilon_{0}}{\epsilon''(\omega_{k})}cos\omega_{k}t .\label{eq:Eq18}
\end{equation}

As typical for most resonances, the steady state is shifted by almost ninety degree in time with respect to the incident field
$\mathbf{E}_{0}(t)$. It is also natural that the magnitude of the steady state is controlled by the ratio of real and imaginary
parts of dielectric permittivity at the resonance frequency. It is also revealing that the resonance magnitude depends on the
spatial orientation of the incident field $\mathbf{E}_{0}(t)$ with respect to the dipole moment $\mathbf{p}_{k}$ of plasmon
resonance mode.

\begin{figure}[h]
\includegraphics[width = 3.4in, height = 2.2in ]{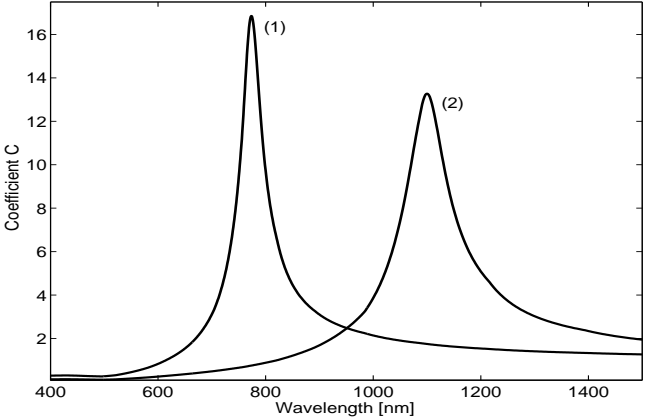}
\caption{\label{fig:fig1} $C(\omega_{0})$ computed for: (1) gold nanocylinders placed on a glass substrate that resonate at
wavelength 774 nm (see \cite{Lamprecht99}) and (2) gold nanorings placed on a glass substrate that resonate at wavelength 1102
nm (see \cite{Aizpurua03}). The Au dispersion relation from \cite{Johnson72} have been used in computations.}
\end{figure}

For off-resonance excitation $f(t)=sin\omega_{0}t$, similar calculations lead to the expression:
\begin{equation}
a_{k}^{(ss)}(t)=(\mathbf{E}_{0}\cdot \mathbf{p}_{k})C(\omega_{0})cos(\omega_{0}t+\varphi),\label{eq:Eq19}
\end{equation}
where
\begin{equation}
C(\omega_{0})= \sqrt{\frac{[\epsilon'(\omega_{0})-\epsilon_{0}]^{2} +
[\epsilon''(\omega_{0})]^{2}}{[\epsilon_{k}-\epsilon'(\omega_{0})]^{2} + [\epsilon''(\omega_{0})]^{2}}}. \label{eq:Eq20}
\end{equation}
Figure 1 illustrates that $C(\omega_{0})$ is narrow peaked at the resonance frequency $\omega_{k}$ for the two cases of plasmon
resonances experimentally studied in \cite{Lamprecht99} and \cite{Aizpurua03}.

\begin{figure}[h]
\includegraphics[width = 3.2in, height = 2in]{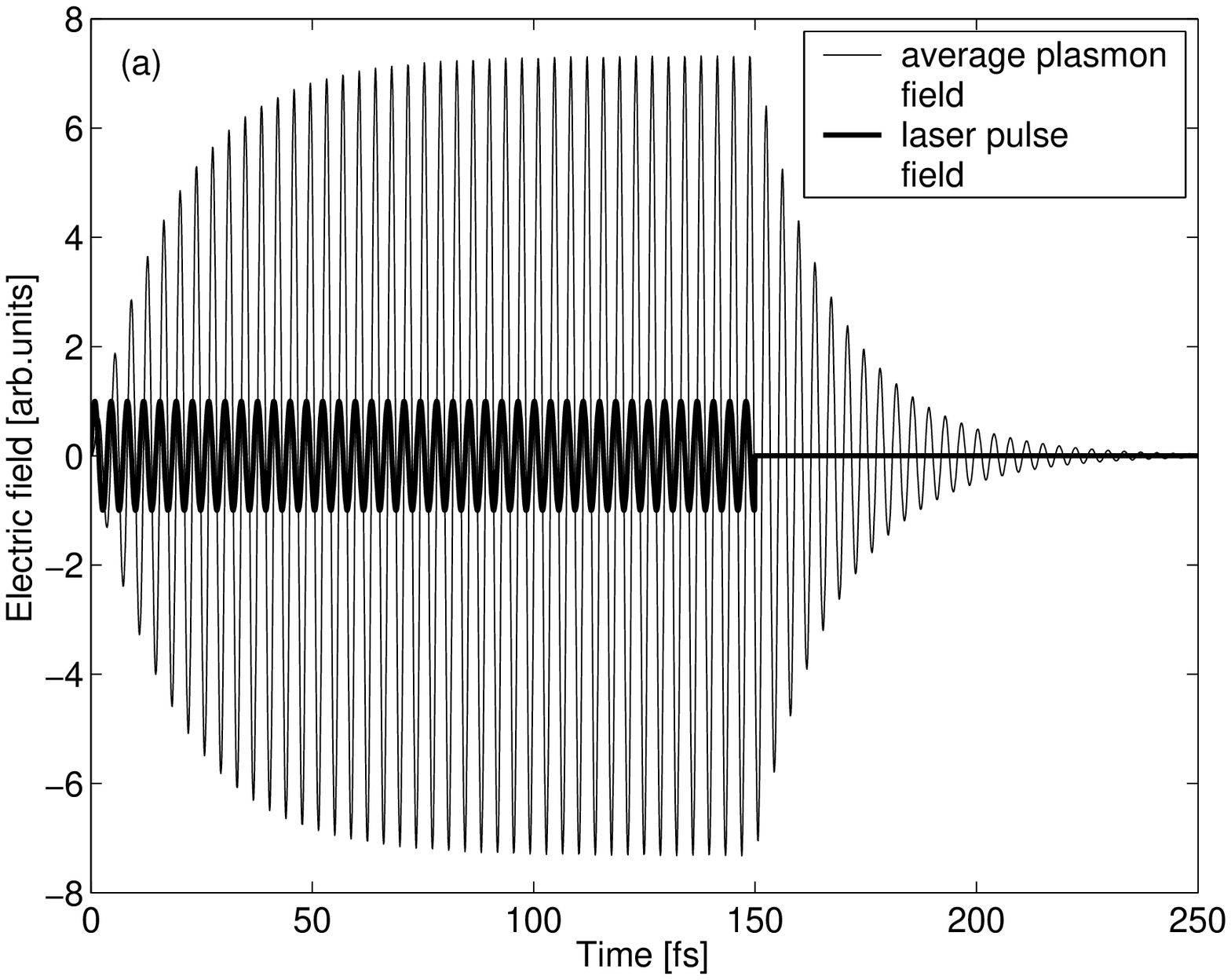}
\hfill
\includegraphics[width = 3.2in, height = 2in]{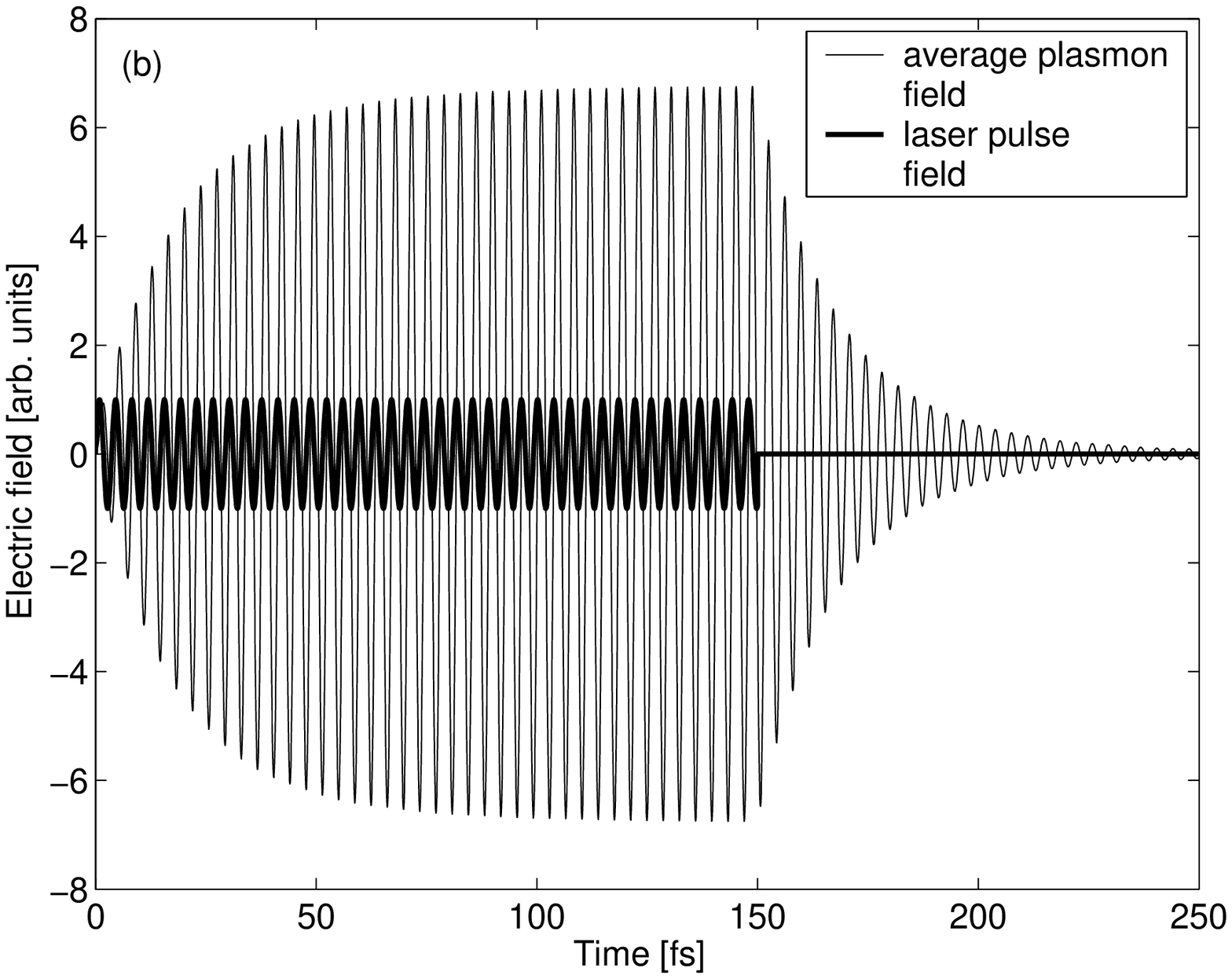}
\caption{ Dynamics of plasmon resonances for Au nanorings on a glass substrate computed for the resonance wavelength 1102 nm
(see \cite{Aizpurua03}) by using (a) Durde model ($\gamma = 1.0753\times 10^{14}~s^{-1}$ \cite{Johnson72}) and (b) Au dispersion
relation from \cite{Johnson72}. } \label{fig:fig2}
\end{figure}

Next, we point out that simple analytical expressions for $g_{k}(t)$ in (\ref{eq:Eq16}) can be obtained for the Drude model of
$\epsilon(\omega)$:
\begin{equation}
\epsilon(\omega)= \epsilon_{0}\left[1-\frac{\omega_{p}^{2}}{\omega (\omega + i\gamma)} \right]. \label{eq:Eq21}
\end{equation}
Indeed, by using (\ref{eq:Eq21}) in (\ref{eq:Eq15}) we obtain:
\begin{equation}
\tilde{g}_{k}(\omega)= -\frac{\omega_{k}^{2}+\gamma^{2}}{(i\omega - \alpha_{1})(i\omega - \alpha_{2})}, \label{eq:Eq22}
\end{equation}
where: $\alpha_{1}=\frac{\gamma}{2}+i\beta$ and $\alpha_{2}=\frac{\gamma}{2}-i\beta$, $\beta=\frac{\sqrt{4 \omega_{k}^{2}+ 3
\gamma^{2}}}{2}$. Now the inverse Fourier transform of (\ref{eq:Eq22}) yields for $t>0$:
\begin{equation}
g_{k}(t)= -\frac{2(\omega_{k}^{2}+\gamma^{2})}{\sqrt{4 \omega_{k}^{2}+ 3 \gamma^{2}}}e^{-\frac{\gamma}{2}t}sin \beta t.
\label{eq:Eq23}
\end{equation}
Formulas (\ref{eq:Eq16}) and (\ref{eq:Eq23}) can be used for analytical calculations of $a_{k}(t)$ for ``rectangular'' laser
pulses $\mathbf{E}^{(0)}f(t)$. Indeed, in the case when:
\begin{equation}
f(t)) =
\begin{cases}
0 & \text{for~~$t<0$,} \\
 sin\omega_{k}t  & \text{for~~$t\geq 0$,}
\end{cases}
\label{eq:Eq24}
\end{equation}
from (\ref{eq:Eq16}), (\ref{eq:Eq23}) and (\ref{eq:Eq24}) we derive:
\begin{equation}
a_{k}(t)=-(\mathbf{E}_{0}\!\cdot \!\mathbf{p}_{k})\omega_{k}e^{-\frac{\gamma}{2}t}\left(\frac{1}{\gamma}cos\beta t
\!-\!\frac{2}{\beta}sin\beta t \right)\! +\! a_{k}^{(ss)}(t), \label{eq:Eq25}
\end{equation}
where
\begin{equation}
a_{k}^{(ss)}(t)= (\mathbf{E}_{0}\cdot \mathbf{p}_{k})\left(\frac{\omega_{k}}{\gamma}cos\omega_{k} t - sin\omega_{k} t \right) .
\label{eq:Eq26}
\end{equation}

In the case of finite rectangular laser pulse
\begin{equation}
f(t)= \begin{cases}
         0 & \text{for~~$t<0$,}  \\
         sin\omega_{k}t & \text{for~~$T\leq t\leq 0$,} \\
         0 & \text{for~~$t>T$,}  \\
       \end{cases}
\label{eq:Eq27}
\end{equation}
from (\ref{eq:Eq16}), (\ref{eq:Eq23}) and (\ref{eq:Eq27}) we derive for $t>T$:
\begin{align}
a_{k}(t)=- &(\mathbf{E}_{0}\cdot
\mathbf{p}_{k})e^{-\frac{\gamma}{2}(t-T)}\Big[\omega_{k}e^{-\frac{\gamma}{2}T}\left(\frac{1}{\gamma}cos\beta t
-\frac{2}{\beta}sin\beta t \right) \nonumber \\
& - \left(\frac{\omega_{k}}{\gamma}cos\omega_{k} t - sin\omega_{k} t \right) \Big]. \label{eq:Eq28}
\end{align}

It is instructive to compare the computational results obtained by using the Drude model with those obtained by using the
experimentally measured dispersion relation $\epsilon(\omega)$. This comparison is presented by Figures 2a and 2b for Au rings
subject to finite rectangular laser pulses. These figures suggest that the Drude model leads to quantitatively similar results
as the use of actual dispersion relation. Formula (\ref{eq:Eq28}) also implies that $1/\gamma$ can be identified as the decay
(dephasing) time for the light intensity of plasmon modes. In accordance with the available data for $\gamma$ (see
\cite{Gay06,Johnson72}), this suggests that formula (\ref{eq:Eq28}) predicts the decay (dephasing) time for plasmon modes in
gold (and silver) nanoparticles in the range of 5-12fs, which is consistent with the experimental results reported in
\cite{Lamprecht99,Lehmann00,Kubo05,Muskens06,Zentgraf04,Gay06,Lamprecht97}.

\begin{figure}[h]
\includegraphics[width = 3.2in, height = 1.9in]{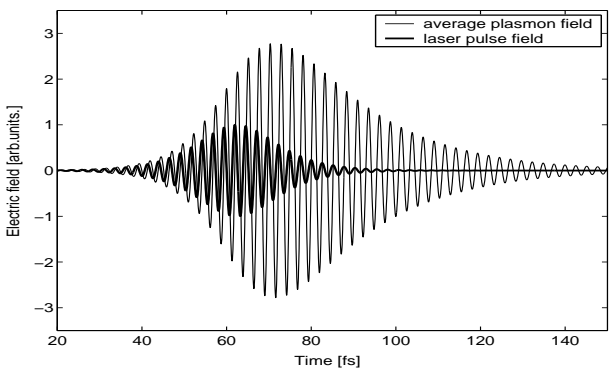}
\caption{\label{fig:fig3} Dynamics of plasmon resonances for Au cylinders on a glass substrate computed for the resonance
wavelength 774 nm (see \cite{Lamprecht99}). The Au dispersion relation from \cite{Johnson72} have been used in computations.}
\end{figure}

Finally, we present the comparison between computational results based on the technique outlined above and experimental results
presented in \cite{Lamprecht99,Zentgraf04}. By using equations (\ref{eq:Eq15})-(\ref{eq:Eq16}) we have computed the
time-dynamics of the 774 nm resonance wavelength plasmon mode for Au cylinders on glass substrate \cite{Lamprecht99}. Figure 3
presents the time variation of the incident electric field of the laser pulse (bold line) used in experiments reported in
\cite{Lamprecht99} and the corresponding computed time-dynamics of the average (over nanoparticle volume) electric field of the
plasmon mode (thin line). The computed plasmon time-dynamics was used to compute the third order autocorrelation function (ACF)
which was compared with the experimentally measured ACF from \cite{Lamprecht99}. The results of comparison presented in Figure
4. Figure 5 presents the comparison between our calculations of the second-order ACF for noncentrosymmetric L-shape gold
nanoparticles from \cite{Zentgraf04} and the experimentally measured in \cite{Zentgraf04} second-order ACF. Figures 4 and 5
suggest the agreement with experimental data within ten percent.

\begin{figure}[h]
\includegraphics[width = 3.2in, height = 2.0in]{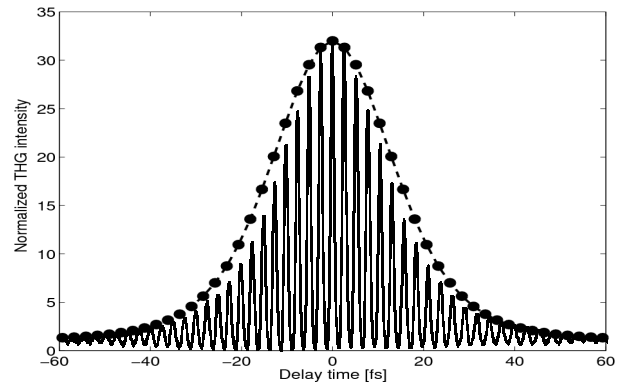}
\caption{\label{fig:fig4a} Envelop of the calculated third order ACF (filled circles) superimpose on the measured third order
ACF (solid line)from \cite{Lamprecht99} for Au cylinders on a glass substrate at wavelength 774 nm (maximum is normalized to
32).}
\end{figure}

\begin{figure}[h]
\includegraphics[width = 3.2in, height = 2.0in]{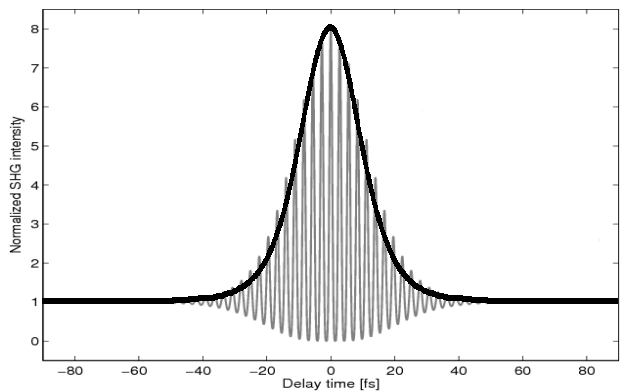}
\caption{\label{fig:fig4b} The calculated second order ACF superimpose on the envelop of measured second order ACF from
\cite{Zentgraf04} for noncentrosymmetric L-shape Au nanoparticles on a tantalum-dioxide substrate at wavelength 838 nm (maximum
is normalized to 8).}
\end{figure}



\end{document}